\documentclass{amsart}[12pt]
\usepackage{amsmath, amssymb}
\usepackage{amscd,amsfonts,amssymb, color}
\usepackage{latexsym, graphicx, pstricks}
\usepackage[pdftex,bookmarks,colorlinks]{hyperref}

\newtheorem{defin}{Definition}
\newtheorem{remark}{Remark}
\newtheorem{theorem}{Theorem}
\newtheorem{lemma}{Lemma}
\newtheorem{corollary}{corollary}
\long\def\symbolfootnote[#1]#2{\begingroup%
\def\thefootnote{\fnsymbol{footnote}}\footnote[#1]{#2}\endgroup}

\newcommand{\sa}{crossover~}
\newcommand{\sae}{crossover~}
\newcommand{\sg}{\Sigma}

\newcommand{\ve}{\varepsilon}

\newcommand{\sub}{\mathrm{sub}}
\newcommand{\rs}{\hspace{-3.5mm}-\hspace{-2mm}<}
\newcommand{\ers}{>\hspace{-2mm}-\hspace{-2mm}<}
\begin{document}
\title{Application of  Generalised sequential crossover of  languages to
 generalised splicing }
\author{L Jeganathan \and R Rama \and Ritabrata Sengupta}
\address{Department of Mathematics\\ Indian Institute of Technology
\\Chennai 600 036, India.}
\email{lj, ramar, rits@iitm.ac.in}
 \maketitle
\begin{abstract}

 This paper  outlines an  application of iterated version of generalised
 sequential crossover of two languages (which in some sense, an
 abstraction of the crossover of chromosomes in living organisms)
 in studying some classes of    the newly proposed generalised
splicing ($GS$) over two languages. It is proved that, for  $X,Y \in
\{FIN, REG, LIN, CF, CS, RE
 \}, \sg \in FIN$,
  the subclass of generalized splicing
 languages namely $GS(X,Y,\sg)$, (which is a subclass of the class
 $GS(X,Y,FIN)$) is always regular.
\end{abstract}

\section{Introduction}
 Tom Head proposed \cite{head} an operation called `splicing', for describing
 the recombination of DNA sequences under the  application of restriction
 enzymes and ligases. Given two strings $u\alpha\beta v$ and $u'\alpha'\beta'
 v'$ over some alphabet  $V$ and a splicing rule
 $\alpha\#\beta\$\alpha'\#\beta'$, two strings $u\alpha\beta' v'$
 and $u'\alpha'\beta v$ are produced. The splicing rule
 $\alpha\#\beta\$\alpha'\#\beta'$ means that the first string is cut
 between $\alpha$ and $\beta$ and the second string is cut between
 $\alpha'$ and $\beta'$, and the fragments recombine crosswise.
 \par The splicing scheme (also written as
 H-scheme) is a pair $\sigma=(V,R)$ where $V$ is an alphabet and
 $R\subseteq V^*\#V^*\$V^*\#V^*$ is the set of splicing rules.
 Starting from a language, we generate a new language by the iterated
 application of splicing rules in $R$. Here $R$ can be infinite.
 Thus $R$ can be considered as a language over $V\cup\{\#,\$\}$.
 Splicing language (language generated by splicing) depends upon the
 class of the language (in the Chomskian hierarchy) to be
 spliced and the type of the splicing rules to be applied. The class
 of splicing language $H(FL_1,FL_2)$ is the set of strings generated
 by  taking any two strings from $FL_1$ and splicing them by the
 strings of $FL_2$. $FL_1$ and $FL_2$ can be any class of languages
 in the Chomskian hierarchy.
 Detailed investigations on computational power of splicing is found in \cite{dna}.

\par Theory of splicing is an abstract model of the recombinant behaviour of the DNAs.
In a splicing system, the two strings that are spliced, are taken
from the same set and the
 splicing rule is from  another set. The reason for taking two strings from the same set
  is, in the DNA recombination, both the objects that are spliced are DNAs.  For example,
  the splicing language in the class $H(FIN,REG)$ is the language generated by taking
  two strings from a finite language and using strings from a regular language as the
   splicing rules. Any general `cut' and `connection' model should include the cutting
   of two strings taken from two different languages. The strings spliced and the splicing
    rules have an effect on the language generated by the splicing process. In short,
    we view a splicing model as having three languages as three components, two strings
    from two languages
    as the first two components, and a splicing rule as the third component.
    We proposed  a generalised splicing model (GS: Generalised
    splicing) in \cite{Je},
     whose splicing scheme is defined as,
\[\sigma(L_1,L_2,L_3):=\{z_1,z_2: (x,y)\models_r (z_1,z_2),~x\in L_1, y\in L_2, r\in L_3\}.\]
\par   Instead of taking two strings from same language, as being done in the
 theory of splicing,  we take  two strings from two different languages. We cut them  by using  rules from a third language.  This means,
taking an arbitrary  word  $w_1(\in L_1)$ and an arbitrary  word
from $w_2(\in L_2)$, we cut them by using an arbitrary rule of
$L_3$. If $L_1=L_2$ in the generalised splicing model, we get the
usual $H$-system.

\par Motivated by the chromosomal crossover in living organisms, an
operation called {\it Generalized sequential crossover (GSCO) of
words and languages} was introduced in the paper \cite{Je1}. The
$GSCO_x$ operation over two strings $u_1xv_1$
  and $u_2xv_2$ overlap at  the substring $x$
   generating the strings $u_1xv_2$ and $u_2xv_1$.

 This $GSCO$  operation differs with the concept of
crossover of the chromosomes in two apects. First,
    in GSCO,
    words of different lengths can participate in a crossover,
    where as homologous chromosomes crossover with each other.
    Second, in $GSCO$ crossing over occurs at only one site between the
    words, whereas chromosomal crossovers can occur at more than one
    site.  Though the $GSCO$ operation cannot be called as the exact abstraction
    of the chromosomal crossover, the study of $GSCO$ over languages
reveals many interesting results such as the iterative GSCO of any
language is always regular.

\par Incidentally, the words generated by the \sa of two strings
over the substring $x$, is the same as the words generated by the
generalised splicing of the strings $u_1xv_1$ and $u_2xv_2$ using
the splicing rule.
 $x\#\$x\#$. The overlapping strings $R$ in the GSCO has a correspondence with the set of
 splicing rules of the generalised splicing model. This correspondence motivated us to
 investigate the generalised splicing for some classes of languages in Chomskian hierarchy.
 \par Though one can develop a theory of generalised splicing on the lines of $H$-system,
 in this paper we investigate a sub-class of the class $GS(X,Y,FIN),~X,Y\in\{FIN,REG,LIN,CF,CS
 ,RE\}$. That is, we investigate the class $(X,Y,R)$ where $R$ is a finite set of words of
 length 1.

In \cite{Je1}, the GSCO of unary languages and its iterated versions
were defined.  For the purpose of our investigation, we define the
 iterated GSCO for two languages $L_1$ and $L_2$   without loosing the sense of the definition given in
\cite{Je1}.

\par Section 2 gives the definition of $GSCO$ of languages along
with some results of \cite{Je1} which are required for our study and
the definition of {\it generalised splicing} as introduced in
\cite{Je}. Section 3 discusses the application of $GSCO$ in studying
some sub classes of generalised splicing.

\section{Preliminaries}

\par Throughout this paper, we assume that the reader is familiar with
the fundamental concepts of formal language theory and automata,
i.e.
 notations finite automata \cite{HMU}.

 \par In this section  we give the formal definition of generalised sequential crossover system
  system as defined in \cite{Je1} along with some results,which are
  required for our investigation.  We also give the formal
  definition of {\it generalised splicing} as in \cite{Je}

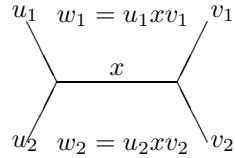
\begin{figure}[h]
\begin{center}
\setlength{\unitlength}{2mm}
\begin{picture}(14,10)(-7,-5)
\put(-4,0){\line(1,0){8}} \put(-6,4){\line(1,-2){2}}
\put(-6,-4){\line(1,2){2}} \put(6,4){\line(-1,-2){2}}
\put(6,-4){\line(-1,2){2}} \put(-7,-4.3){$u_2$} \put(-7,4.3){$u_1$}
\put(6.2,-4.3){$v_2$} \put(6.2,4.3){$v_1$} \put(-0.5,0.5){$x$}
\put(-4,4){$w_1=u_1xv_1$} \put(-4,-4.5){$w_2=u_2xv_2$}
\end{picture}
\end{center}
\caption{A scheme for \sa of two strings} \label{eqi}
\end{figure}

 \begin{defin}
Generalised sequential \sa scheme $GSCO=(\sg,R)$, where $\sg$ is the
finite alphabet,
 $R\subseteq \sg^*$ be the set of overlapping strings; we write $GSCO=(\sg,R)$
 as $GSCO_R$. $GSCO_R$ is also called a $R$-crossover. When $R$ is singleton,
  say $R=\{x\}$, we write $GSCO_x$ instead of $GSCO_R$.
%$>^x\hspace{-3.5mm}-\hspace{-2mm}<$

\par For a given GSCO scheme $GSCO$ and two words $w_1=u_1xv_1$ and $w_2=u_2xv_2\in \sg^*$,
 we define
\[GSCO_x(w_1,w_2)=\{u_1xv_2,u_2xv_1\in\sg^*:  w_1=u_1xv_1, w_2=u_2xv_2,~
\ve\neq x\in R\}.\] The scheme is shown in figure \ref{eqi}.
\end{defin}

\par Instead of writing $GSCO_x(u_1xv_1, u_2xv_2)$, we also write
 $u_1xv_1>^x\rs u_2xv_2=\{ u_1xv_2,u_2xv_1\}$, which means that the two strings
  $u_1xv_1$ and $u_2xv_2$ \sae over the sub-string $x$ to generate two new
  words  $u_1xv_2$ and $u_2xv_1$. We also write $u_1xv_1>^x\rs u_2xv_2=\{u_1xv_2,u_2xv_1\}$
   instead of $(u_1xv_1,u_2xv_2)>^x\rs\{u_1xv_2,u_2xv_1\}$. Then
\[GSCO_R(w_1,w_2)=\bigcup_{x\in R}w_1>^x\rs w_2.\]
  Obviously $R$ should contain words which are sub-words found in both
  $w_1$ and $w_2$, otherwise $GSCO_R(w_1,w_2)$ will be empty. We call the operation
  $GSCO_x,~x\in\sg$ as the symbol overlapping GSCO. Similarly we call $GSCO_x,~x\in\sg^*$
  as the string overlapping GSCO. Let $\sub(w)$ be the set of all sub-words of $w$.
   If in a GSCO scheme $R=\sub(w_1)\cap\sub(w_2)$, we simply write $GSCO(w_1,w_2)$, i.e.
    $GSCO(w_1,w_2)$ is the set of all words that can be generated by the GSCO
     of $w_1$ and $w_2$ with all possible overlapping. In other words,
\[GSCO(w_1,w_2)= \bigcup_x GSCO_x(w_1,w_2),~~~~x\in\sub(w_1)\cap\sub(w_2).\]

\par We extend the above definition  to languages. Given any two languages $L_1$ and $L_2$ over the alphabet $\sg_1$ and $\sg_2$ respectively such that $\sg_1\cap\sg_2\neq\emptyset$, we define
\[GSCO_R(L_1,L_2)=\bigcup_{\substack{w_1\in L_1\\w_2\in L_2}}GSCO_R(w_1,w_2).\]
Here the underlying crossover scheme is $GSCO=(\sg_1\cup\sg_2,R)$.
As mentioned earlier, when  $R=\sub(L_1)\cap\sub(L_2)$ ($R$ is the
set of all possible overlapping between a word of \(L_1\) and a word
of \(L_2\)).
\[GSCO_R(L_1,L_2)=\bigcup_{\substack{w_1\in L_1\\w_2\in L_2}}GSCO(w_1,w_2).\]
\(GSCO(L,L)\) is written as just \(GSCO(L)\).
\par In computing $GSCO (w_1,w_2)$, one has to first compute all the common sub-strings $x$
and compute $\bigcup_xGSCO_x(w_1,$ $w_2)$. For $GSCO(L)$  we have to
compute $\bigcup_{w_1,w_2\in L}GSCO(w_1,w_2)$. In short,
\[GSCO(L)=  \bigcup_{w_1,w_2\in L}~\bigcup_xGSCO(w_1,w_2),\hspace{3mm}x\in
\sub(w_1)\cap\sub(w_2) ,\] which increases the complexity of the
computation of GSCO. We have the following theorem to reduce this
tedious calculation of finding all the common sub-strings of all the
pairs of  words of a given language $L$.
\begin{theorem}
Let $w_1,w_2\in\sg^*$.
\[GSCO(w_1,w_2)=\bigcup_{a\in\sg_{w_1}\cap\sg_{w_2}}GSCO_a(w_1,w_2).\]
\end{theorem}
\begin{corollary}\label{c1}
$GSCO(w_1,w_2)=\bigcup_{a\in\sg}GSCO_a(w_1,w_2)$.
\end{corollary}
\begin{corollary}
$GSCO(L)=\bigcup_{w_1,w_2\in L}\bigcup_{a\in\sg}GSCO_a(w_1,w_2).$
\end{corollary}
This corollary tells us that to compute $GSCO(L)$ it is enough to
compute the GSCO of
 $w_1$ and $w_2$ over the symbols of the alphabet $\sg$ and take the union of all those
 $ GSCO(w_1,w_2)$'s.
\par The operation GSCO is called 1-GSCO if in all the concerned overlapping,
 we consider the word which has the prefix of the first word and the suffix of the
  second word as the only word generated. So $1GSCO_x(u_1xv_1,u_2xv_2)=\{u_1xv_2\}$,
  i.e. the operation 1GSCO generates only one word. We denote 1GSCO by $>_1\rs$.
 The operation GSCO is called 2GSCO if in all the concerned
overlapping we consider
 both the words generated. So the operation 2GSCO coincides with
 GSCO. It is proved that $1GSCO(L)$ is equal to $GSCO(L)$

  We define two types of iterated GSCO namely,
unrestricted iterative closure of GSCO and the restricted iterative
closure of GSCO.
\begin{defin}
Given a language $L$, we define the language obtained from $L$ by
unrestricted iterated application of GSCO. This language, called the
unrestricted GSCO closure of $L$, denoted by $uGSCO^*(L)$, is
defined as
\begin{eqnarray*}
uGSCO^0(L)&=&L\\
uGSCO^{i+1}(L)&=&uGSCO^{i}(L)\cup uGSCO(uGSCO^i(L))\\
uGSCO^*(L)&=&\bigcup_{i\geq0}uGSCO^{i}(L)
\end{eqnarray*}
\end{defin}

\begin{defin}
The restricted closure of GSCO denoted by $rGSCO^*(L)$ is defined
recursively as follows:-
\begin{eqnarray*}
rGSCO^0(L)&=&L\\
rGSCO^{i+1}(L)&=&rGSCO(rGSCO^i(L),L)\hspace{0.5in}i\geq1\\
rGSCO^*(L)&=&\bigcup_{i\geq0}rGSCO^i(L)
\end{eqnarray*}
\end{defin}

\begin{theorem}
$r1GSCO^*(L)=u1GSCO^*(L)$.
\end{theorem}

\begin{theorem}
For a language $L$, $GSCO^*(L)$  is a regular language.
\end{theorem}
We give the defintion of generalised splicing model.
\begin{defin}[Generalised splicing scheme]\label{main}
 Generalised splicing scheme is defined as a 2-tuple
  $\sigma=(\sg,R)$, where $\sg$ is an alphabet, and
  $R\subseteq \sg^*\#\sg^*\$\sg^*\#\sg^*$. Here $R$ can be infinite, and $R$
  is considered as a set of strings, hence a language. For a given $\sigma$,
  and  languages $L_1\subseteq \sg^*$ and $L_2\subseteq \sg^*$, we define
\[\sigma(L_1,L_2,R)=\{z_1,z_2:(x,y)\models_r (z_1,z_2),~\mbox{for~} x\in L_1, y\in L_2,
 r\in R\}.\]
We refer the generalised splicing scheme $\sigma =(\sg, R)$ as
$\sigma_R. $
\end{defin}
\section{Application of GSCO to generalised splicing}
%\input{spintro.tex}

%\begin{defin}[Generalised splicing scheme]\label{main}
% Generalised splicing scheme is defined as a triplet
%  $\sigma=(\sg,R)$, where $\sg$ is an alphabet, and
%  $R\subseteq \sg^*\#\sg^*\$\sg^*\#\sg^*$. Here $R$ can be infinite, and $R$
%  is considered as a set of strings, hence a language. For a given $\sigma$,
%  and  languages $L_1\subseteq \sg^*$ and $L_2\subseteq \sg^*$, we define
%\[\sigma(L_1,L_2,R)=\{z_1,z_2:(x,y)\models_r (z_1,z_2),~\mbox{for~} x\in L_1, y\in L_2,
% r\in R\}.\]
%We refer the generalised splicing scheme $\sigma =(\sg, R)$ as
%$\sigma_R. $
%\end{defin}
We give  the definition of  an  iterative  $GSCO$ over two languages
as well as give the iterated version of the generalised splicing for
our purpose of investigation.
\begin{defin}
  Let
$L_1$ and $L_2$ be any two languages. The iterated $GSCO$ of
$L_1$and $L_2$ is defined as follows.
\begin{eqnarray*}
GSCO^0(L_1,L_2) &= &(L_1 \cup L_2);\\
 GSCO^1(L_1,L_2)&=&\cup_{w_1 \in L_1,w_2 \in L_2}GSCO^1(w_1,w_2);\\
GSCO^{i+1}(L_1,L_2)&=&GSCO^i(L_1,L_2)~\cup~
GSCO(GSCO^i(L_1),GSCO^i(L_2))\\
GSCO^*(L_1,L_2)&=& \cup_{i \geq 0}GSCO^i(L_1,L_2)\\
\end{eqnarray*}
\end{defin}
$GSCO^*(L,L)$ is written as $GSCO(L)$ itself.  The above definition
of $iterated ~ GSCO$ over two languages   is more logical in the
sense that, when $L_1~=~L_2$, the above iterative definition reduces
to $GSCO^*(L)$

We define the iterated generalized splicing as follows. Let
$\sigma_R~=~(\sg,R)$ be the generalised splicing scheme.
\begin{defin}
  Let
$L_1$ and $L_2$ be any two languages.  For a given generalised
splicing scheme $ \sigma_R$,the iterated generalised splicing of
$L_1$and $L_2$ is defined as follows.
\begin{eqnarray*}
\sigma_R^0(L_1,L_2)&=& L_1 \cup L_2\\
\sigma_R^1(L_1,L_2)&=&\{z | (w_1,w_2) \vdash_r z, ~w_1 \in L_1,~ w_2
\in
L_2, ~r \in R\}\\
\sigma_R^2(L_1,L_2)&=&\sigma_R^1(L_1,L_2)\cup\sigma_R(\sigma_R^1(L_1),
\sigma_R^1(L_2))\\
\sigma_R^{i+1}(L_1,L_2)&=&\sigma_R^i(L_1,L_2)\cup\sigma_R(\sigma_R^i(L_1),
\sigma_R^i(L_2))\\
\sigma_R^*(L_1,L_2)&=& \cup_{i \geq 0} \sigma_R^i(L_1,L_2)
\end{eqnarray*}
The language of the generalised splicing of $L_1,L_2$ and $R$ is
defined as
\[ GS(L_1,L_2,R)= \sigma_R^*(L_1,L_2) \]
\end{defin}
$\sigma_R(L,L)$ is written as $\sigma_R(L)$.

We have the following lemma whose proof is immediate.
\begin{lemma}
For any two languages $L_1,L_2$,
$GSCO_R(L_1,L_2)=\sigma_R(L_1,L_2)$; For any langauge $L$,
$GSCO_R(L)=\sigma_R(L)$
\end{lemma}

\begin{theorem}
For any language $L$,  $GSCO_R^i(L)~=~\sigma_R^i(L)$, for
$i~\geq~0$.
\end{theorem}
\begin{proof}
We prove by the method of induction on $i$. When $i=0$, the result
is trivially true.  When $i=1$ also, it is true.  Assume that the
result is true for $i=0,1,2...n$.  We have
$GSCO_R^n(L)=\sigma_R^n(L)$.
$GSCO_R^{n+1}(L)=GSCO(GSCO_R^n(L))~\cup~
GSCO_R^n(L)=\sigma_R(\sigma_R^n(L))
\cup\sigma_R^n(L)=\sigma_R^{n+1}(L)$.  Hence the proof.
\end{proof}
\begin{theorem}
Let $L_1$ and $L_2$ be any two languages.  Let $V_{L_1}$ and
$V_{L_2}$ be the alphabets of $L_1$and $L_2$ respectively.  Let
$R=V_{L_1}\cap V_{L_2}$. Then
\[GSCO^*(L_1,L_2)=GS(L_1,L_2,R)\]
\end{theorem}
\begin{proof}
We claim $GSCO_R^i(L_1,L_2)=\sigma_R^i(L_1,L_2),~i\geq0$. We follow
the method of induction. For, $i=0,~1$ it is true.  Assume that the
result is true for $i=2,3,...,n$.  Consider
$GSCO_R^{n+1}(L_1,L_2)=GSCO(GSCO_R^{n}(L_1), GSCO_R^{n}(L_2))\\\cup
GSCO_R^n(L_1,L_2) =GSCO(\sigma_R^n(L_1),~\sigma_R^n(L_2))\cup
\sigma_R^n(L_1,L_2)$ (by induction)
$=\sigma(\sigma_R^n(L_1),\sigma_R^n(L_2))\cup
\sigma_R^n(L_1,L_2)=\\~\sigma_R^{n+1}(L_1,L_2).$ Consider
$GSCO_R^{i}(L_1,L_2)=GSCO(GSCO_R^{i-1}(L_1), ~GSCO_R^{i-1}(L_2))\cup
GSCO_R^{i-1}(L_1,L_2)=$ $\sigma(\sigma_R^{i-1} (L_1),$
$\sigma_R^{i-1}(L_2))
\cup\sigma_R^{i-1}(L_1,L_2)=\sigma_R^i(L_1,L_2)$.  Thus we have,
$GSCO_R^i(L_1,L_2)=\sigma_R^i(L_1,L_2), $for every$~ i \geq 0.$ This
implies, $GSCO^*(L_1,L_2)=\sigma_R^*(L_1,L_2)~=$ $~GS(L_1,L_2,R)$
\end{proof}

\begin{theorem}
For any two languages $L_1$ and $L_2$,
$GSCO^*(L_1,L_2)=GSCO(GSCO^*(L_1),GSCO^*(L_2))$
\end{theorem}
\begin{proof}
Let $w\in GSCO^*(L_1,L_2)$. This implies $w\in GSCO^i(L_1,L_2)$, for
some $i$. Then, there exist $w_1\in GSCO^{i-1}(L_1),$ $w_2\in
GSCO^{i-1}(L_2)$ such that $w\in
 GSCO(GSCO^{i-1}(L_1),GSCO^{i-1}(L_2))\Rightarrow w\in
GSCO(GSCO^*(L_1),$ $GSCO^*(L_2))\Rightarrow GSCO^*(L_1,L_2)\subseteq
GSCO(GSCO^*(L_1), GSCO^*(L_2))$.
\par Let $w\in GSCO(GSCO^*(L_1),GSCO^*(L_2))$.  That is, there exist
$w_1\in GSCO^*(L_1)$ and $w_2\in GSCO^*(L_2)~$such that$w\in
GSCO(w_1,w_2)$.  Without any loss of any generality, we suppose that
$w_1\in GSCO^i(L_1)$ and $w_2\in GSCO^j(L_2)~$ such that $i<j$.
Since $GSCO^i(L_1)\subseteq GSCO^j(L_1)$, we have, $w_1\in
GSCO^j(L_1)$ and $w_2\in GSCO^j(L_2)~$. That is, $w\in
GSCO(GSCO^j(L_1),GSCO^j(L_2))$. This implies, $w \in
GSCO(GSCO^j(L_1),GSCO^j(L_2)) \cup GSCO^j(L_1,L_2)\Rightarrow w\in
GSCO^*(L_1,L_2)\Rightarrow$ $GSCO(GSCO^*(L_1),GSCO^*(L_2))\subseteq
GSCO^*(L_1,L_2)$.  Hence the proof.
\end{proof}
\begin{corollary}
$GS(L_1,L_2,R)~=~GSCO_R(GSCO_R^*(L_1),GSCO_R^*(L_2))$
\end{corollary}
\begin{remark}
The above corollary can be used to compute $GS(L_1,L_2,R)$
\end{remark}
\begin{theorem}\label{l1l2reg}
Let $L_1$ and $L_2$ be any two regular languages.  Then,
$GSCO(L_1,L_2)$is regular.
\end{theorem}
\begin{proof}
Here $L_1,L_2$ are two regular languages.  We  have two finite
automata $M_1~=~(Q_1,V_1,\delta_1,q_1,f_1)$ ,
$M_2~=~(Q_2,V_2,\delta_2,q_2,f_2)$ such that $L(M_1)=L_1$ and
$L(M_2)=L_2$ respectively.  If $V_1 \cap V_2$ is empty, the result
is trivial.  Assume that $\emptyset \neq V_1 \cap V_2
=\{a_1,a_2,\cdots, a_n\} $. We group the transitions of $\delta_1$
as $ \delta_{1,a_i}$, for every $i,~ 1 ~\leq i ~\leq n$ as follows.
$\delta_{1,a_i}$ is the  set of all transitions of $\delta_1$ of the
form $\delta_1(p,a_i)=q,$ where $p,q \in Q_1, a_i \in V_1 \cap V_2$.
That is $\delta_{1,a_i}$ is the set of all transitions of $M_1$
which corresponds to an edge with label $a_i$ in the transition
graph of $M_1$.  We order the transitions in $\delta_{1,a_i}$ in any
way. We call the first transition in $\delta_{1,a_i}$ as
$\delta^1_{1,a_i}$ and the second transition as $\delta^2_{1,a_i}$
and so on. A transition of the form $\delta_1(p,a_1)=q, p,q \in Q_1,
a_1 \in V_1 \cap V_2$ will be referred as $\delta^j_{1,a_i},$ for
some $j$. Similarly, we compute the set $\delta_{2,a_i}$ for every
$a_i \in V_1 \cap V_2$ and identify the transitions
$\delta^j_{2,a_i},  $ for some $j  $.\\

We construct finite automata, $B_{i,j,a_k}$, for $1 \leq k \leq n $
and for all possible $i$ and $j$, as follows.
 {\it Construction of FA} : $B_{i,j,a_k}$:
 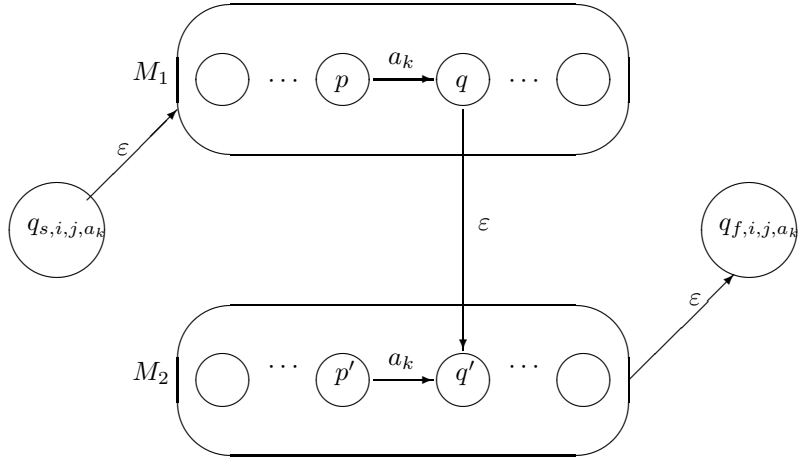
\begin{figure}[h]

\setlength{\unitlength}{2mm}
\begin{center}
\begin{picture}(50,30)(-25,-15)
\put(0,10){\oval(30,10)} \put(0,-10){\oval(30,10)}
\multiput(-12,10)(8,0){4}{\circle{3.5}} \put(-4.5,9.5){$p$}
\put(3.5,9.5){$q$} \put(-2,10){\vector(1,0){4}}\put(-1,11){$a_k$}
\put(-9,9.5){$\cdots$} \put(7,9.5){$\cdots$}

\multiput(-12,-10)(8,0){4}{\circle{3.5}} \put(-4.5,-10){$p'$}
\put(3.5,-10){$q'$} \put(-2,-10){\vector(1,0){4}}\put(-1,-9){$a_k$}
\put(-9,-9.5){$\cdots$} \put(7,-9.5){$\cdots$}

\put(4,8){\vector(0,-1){16}}\put(5,0){$\varepsilon$}

\put(-23,0){\circle{6}}\put(-25,0){$q_{s,i,j,a_k}$}

\put(23,0){\circle{6}}\put(21,0){$q_{f,i,j,a_k}$}

\put(-21,2){\vector(1,1){6}}

\put(15,-10){\vector(1,1){7}}

\put(-19,5){$\varepsilon$} \put(19,-5){$\varepsilon$}

\put(-18,10){$M_1$}

\put(-18,-10){$M_2$}
\end{picture}
\end{center}
\caption{Model of the automata : $B_{i,j,a_k}$}
\end{figure}

\par We define automata $B_{i,j,a_k}$ for every possible $i$ and $j$.\\
 $B_{i,j,a_k}=( Q_1 \cup Q_2 \cup \{q_{s,i,j,a_k},
q_{f,i,j,a_k}\}_{i,j}$ \footnote{By $\{q_{s,i,j,a_k},
q_{f,i,j,a_k}\}_{i,j}$, we mean the set $\{q_{s,1,1,a_k},
q_{f,1,1,a_k} \} \cup \{ q_{s,1,2,a_k},q_{f,1,2,a_k}\} \cup \cdots ~
\cup \{ q_{s,i,j,a_k},q_{f,i,j,a_k}\cdots\}$.  It is clear that
$\{q_{s,i,j,a_k}, q_{f,i,j,a_k}\}_{i,j}$ is finite},

 $\Sigma_1 \cup
\Sigma_2 \cup \{ \ve \},
 \delta_{i,j,a_k}, q_{s,i,j,a_k}, f_2)$ where $ \delta_{i,j,a_k}$ is
 defined as follows.
 \begin{enumerate}
 \item $ \delta_{i,j,a_k}(q_{s,i,j,a_k}, \ve)= q_1$
 \item All the transitions of $\delta_1$
 \item All the transitions of $\delta_2$
 \item If $\delta^i_{1,k}$ is the transition $\delta_1(p,a_k)=q, for some p,q \in
 Q_1 $and if $\delta^j_{2,k}$ is the transition $\delta_2(p',a_k)=q',$ for some $p',q' \in
 Q_2$, include $\delta_{i,j,a_k}(q,\ve)=q'$
 \item $\delta_{i,j,a_k}(q_2,\ve)=q_{f,i,j,a_k}$
 \end{enumerate}
 $B_{i,j,a_k}$ for every possible $i,j$, stands for the
collection of automatons viz., $B_{1,1,a_k},B_{1,2,a_k},\cdots ,\\
B_{2,1,a_k},B_{2,2,a_k},\cdots , \cdots,
B_{i,1,a_k},B_{i,2,a_k},\cdots, $. We have the result
\[GSCO(L_1,L_2)=\cup_{w_1 \in L_1,w_2 \in L_2}GSCO(w_1,w_2)=
\cup_{w_1 \in L_1,w_2 \in L_2}\cup_{a\in\sg_1 \cap
\sg_2}GSCO_a(w_1,w_2).\]  If $w_1 \in L_1, w_2 \in L_2$ has a common
symbol $a_k$ (which may occur more than once in $w_1$ and $w_2$) in
them, $w_1$ and $w_2$ could crossover at $a_k$.  The first
occurrence of $a_k$ in $w_1$ may crossover with the first occurrence
of $w_2$ or the  first occurrence of $a_k$ in $w_1$ may crossover
with the second occurrence of $w_2$ and so on. If the word $w_1$ and
$w_2$ crossover at the symbol $a_k$ such that the $first$ occurrence
of $a_k$ in $w_1$ overlaps with the $second $ occurrence of $a_k$ in
$w_2$ generating  a word $w$,  then $w$ will be accepted by the
automaton $B_{1,2,a_k}$.  We claim that the union of the  languages
accepted by the automata $B_{i,j,a_k}$, for every possible $i,j$, is
$GSCO_{a_k}(L_1,L_2)$\\

{\bf Claim : $\bigcup_{i,j} L(B_{i,j,a_k}) = GSCO_{a_k}(L_1,L_2)$}\\

\begin{eqnarray*}
Suppose ~~w \in \bigcup_{i,j} L(B_{i,j,a_k})
\Rightarrow &  w \in   L(B_{i,j,a_k})\\
\Rightarrow &  w \in Pref(L_1).a_k.Suff(L_2), a_k \in V_1 \cap V_2 \\
\Rightarrow  & w \in  Pref(w_1).a_k.Suff(w_2), w_1 \in L_1, w_2 \in L_2  \\
\Rightarrow  &  w  \in w_1 \ers w_2  \\
\Rightarrow & w \in GSCO_{a_k}(w_1,w_2), w_1 \in L_1, w_2 \in L_2  \\
\end{eqnarray*}

Hence, $\bigcup_{i,j} L(B_{i,j,a_k}) \subseteq GSCO_{a_k}(L_1,L_2).$\\
For the other way, suppose $w \in GSCO_{a_k}(L_1,L_2)$. Then, there
exists $ w_1 \in L_1, w_2 \in L_2$ such that $ w \in
GSCO_{a_k}(w_1,w_2)$. $a_k$ occurs in both $w_1$ and $w_2$. That is,
$w_1 = u_1 a_k u_2 ; w_2 =v_1 a_k v_2$, for some $u_1, u_2,v_1$ and
$v_2$.
  We have the  {\it accepting configuration sequence} for
   $ w_1 \in M_1$ as $q_1u_1pa_kqu_2f_1$ and
  an {\it accepting configuration sequence}
   for $w_2 \in M_2$ as $q_1u_1pa_kqu_2f_1$.  This implies
  that there is an {\it accepting configuration sequence} $q_{s,i,j,a_k} \ve q_1u_1pa_kq'v_2f_2
  \ve q_{f,i,j,a_k}$ in $B_{i,j,a_k}$ such that $u_1a_kv_2 \in
  L(B_{i,j,a_k})$, for some $i,j$.  In the sequence for $M_1$, $a_k$
  can occur more than once. Similar is the case with $M_2$.  Hence,
  $u_1a_kv_2 \in \bigcup_
  {i,j}
  L(B_{i,j,a_k})$. That is, $w \in \bigcup_{i,j}
  L(B_{i,j,a_k})$.  Thus, $ GSCO_{a_k}(L_1,L_2) \subseteq \bigcup_{i,j}
  L(B_{i,j,a_k})$, which proves our claim.

So far, we have constructed an automata which will accept the
$GSCO_{a_k}(L_1,L_2)$, for a given $a_k$.  For $GSCO(L_1,L_2)$, we
have to consider the union of  all such  $GSCO_{a_k}(L_1,L_2)$'s.
So, we  construct an automaton whose language will be the union of
the languages accepted by the automata $B_{i,j,a_k} $, which will
ultimately accept the language $GSCO(L_1,L_2)$.  We construct an
automaton $M= (\{q_s,q_f\} \cup \{q_{s,i,j,a_k},
q_{f,i,j,a_k}\}_{i,j,k} \cup Q_1 \cup Q_2, \Sigma_1 \cup \Sigma_2
\cup \{ \ve \}, \delta, q_s, q_f)$, where $\delta$ is defined as
follows.
\begin{enumerate}
\item $\delta(q_s, \ve)=\{q_{s,i,j,a_k}\}$ for every $i,j$ and $a_k
\in V_1 \cap V_2.$
\item All the transitions of $ \delta_{i,j,a_k}$, for every $i,j$ and
$a_k$.
\item $\delta(q_{f,i,j,a_k}, \ve)= q_f$.
\end{enumerate}

\begin{figure}[h]

\setlength{\unitlength}{1.6mm}
\begin{center}
\begin{picture}(70,70)(-35,-20)
\multiput(-15,30)(0,-25){3}{\framebox(30,20)}

\put(-12,44){\framebox(24,4){$B_{1,1,a_{1}}$}}
\put(-12,34){\framebox(24,4){$B_{i,j,a_{1}}$}}

\multiput(0,40)(0,-9){2}{$\vdots$}

\put(-12,19){\framebox(24,4){$B_{1,1,a_{2}}$}}
\put(-12,9){\framebox(24,4){$B_{i,j,a_{2}}$}}

\multiput(0,15)(0,-9){2}{$\vdots$}

\put(0,1){$\vdots$} \put(-12,-6){\framebox(24,4){$B_{1,1,a_{n}}$}}
\put(-12,-16){\framebox(24,4){$B_{i,j,a_{n}}$}}

\multiput(0,-10)(0,-9){2}{$\vdots$}

\multiput(-30,15)(60,0){2}{\circle{10}} \put(-31,15){{\Large $q_s$}}
\put(29,15){{\Large $q_f$}}

%\qbezier(-27,17)(-18,35)(-12,46)
%\qbezier(12,46)(18,35)(27,17)      %first lines

%\qbezier(-27,16.8)(-18,35)(-12,36)
\put(-27,16.8){\vector(3,4){15}}

\put(12,36){\vector(3,-4){15}}      %second line

\put(-27,16.3){\vector(3,1){15}}
\put(12,20){\vector(3,-1){15}}      %third line

\put(-27,16.3){\vector(1,0){15}}\put(-18,17){$\varepsilon$}

\put(-27,16.3){\vector(3,-1){15}}
\put(12,10){\vector(3,1){15}}       %fourth line

\put(10,15){\vector(1,0){15}}\put(18,15.5){$\varepsilon$}

\put(-27,15){\vector(3,-4){15}}

\put(12,-4){\vector(3,4){14}}       %fifth line

%\put(-29,12){\vector(3,-4){18}}

%\put(-31,18){\vector(3,4){20}}

\put(-28,18){\line(0,1){20}}\put(-28,38){\vector(2,1){15}}
\put(28,38){\vector(0,-1){20}}\put(12,46){\line(2,-1){16.1}}   %top
\put(-27.5,17.5){\vector(2,3){15}}\put(12,40){\vector(2,-3){15}}

\put(-28,12){\line(0,-1){20}}\put(-28,-8){\vector(2,-1){15}}
\put(28,-8){\vector(0,1){20}}\put(12,-16){\line(2,1){16.1}}     %bottom
\put(-27.5,12){\vector(2,-3){15}}\put(12,-10){\vector(2,3){15}}

\multiput(-23,42)(44,0){2}{$\varepsilon$}
\multiput(-21,30)(40,1){2}{$\varepsilon$}
\multiput(-18,27)(35,-1){2}{$\varepsilon$}

\multiput(-20,20)(38,-1){2}{$\varepsilon$}
\multiput(-20,12)(38,-1){2}{$\varepsilon$}

\multiput(-20,7)(38,0){2}{$\varepsilon$}
\multiput(-18,-2)(37,0){2}{$\varepsilon$}
\multiput(-20,-11)(40,0){2}{$\varepsilon$}

\end{picture}
\end{center}
\caption{Model of the automaton M}
\end{figure}
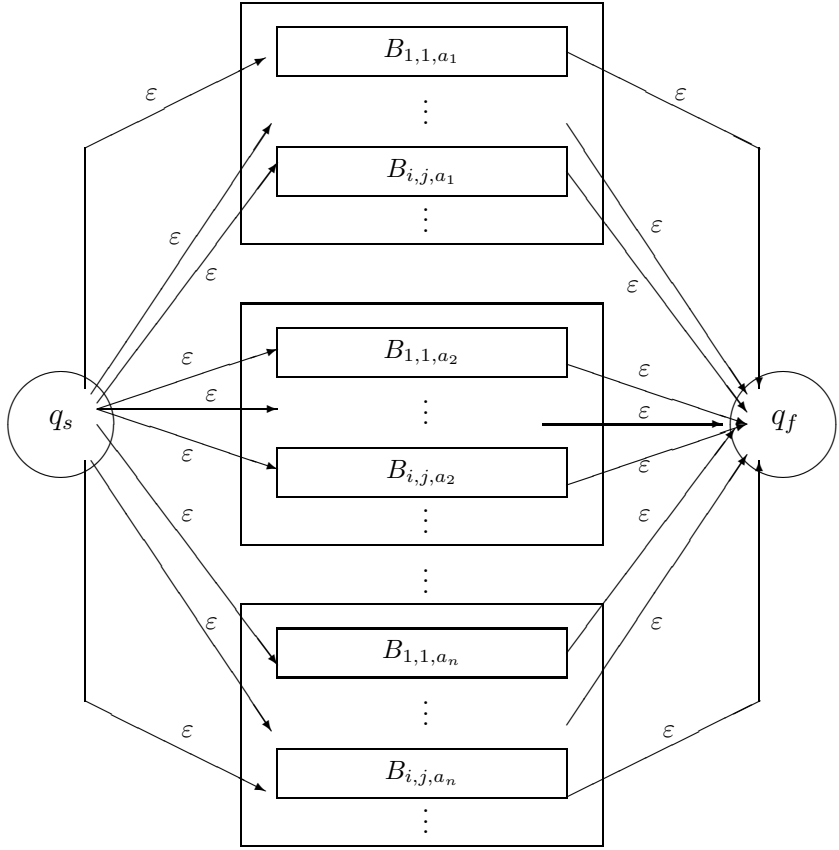

We claim $L(M)=GSCO(L_1,L_2)$. Let $w \in L(M)$. That is, $w \in
\bigcup_{a_k \in V_1 \cap V_2}\bigcup_{i,j} L(B_{i,j,a_k})$. $ w \in
\bigcup_{a_k}GSCO_{a_k}(L_1,L_2)$. This implies $w \in
GSCO(L_1,L_2)$.
Hence $L(M) \subseteq GSCO(L_1,L_2)$. \\
For the other way, let $ w \in GSCO(L_1,L_2)$. Then, $w \in
\bigcup_{a_k} GSCO_{a_k}(L_1,L_2)$.  That is $w \in \bigcup_{a_k}
\bigcup_{i,j}L(B_{i,j,a_k})$. Thus, $w \in L(M)$, which implies
$GSCO(L_1,L_2) \subseteq L(M)$.  Thus, we have constructed a finite
automaton $M$ which accepts $GSCO(L_1,L_2)$.  Hence, $GSCO(L_1,L_2)$
is regular.
\end{proof}
\begin{theorem}\label{reg*l1l2}
Let $L_1$ and $L_2$ be any two languages.  Then, $GSCO^*(L_1,L_2)$
is also regular.
\end{theorem}
\begin{proof}
We have $GSCO^*(L_1,L_2) = GSCO(GSCO^*(L_1),GSCO^*(L_2))$.
$GSCO^*(L_1)$ and $GSCO^*(L_2)$ are regular. By theorem
\ref{l1l2reg}, $GSCO^*(L_1,L_2)$ is regular for any $L_1$ and $L_2$.
\end{proof}
\par Though one can investigate the theory of generalised splicing on
the lines of $H-System$, we attempt to investigate a subclass of the
class of  generalised splicing languages $GS(X,Y,FIN)$, where $X,Y
\in \{FIN, REG, LIN,$ $ CF, CS, RE \}.$  That is, we investigate the
sub class $GS(X,Y,R)$, where $R$ is the  alphabet of the languages
in the class $X$ and $Y$, which is finite. The following theorem
tells that $GSCO^*(X,Y,R)$ are always regular.

\begin{theorem}
$GS(X,Y,R)$, where $X,Y \in \{FIN, REG, LIN, CF, CS, RE \}, R $ is
the common symbols of the languages in $X$ and $Y$,  is regular.
\end{theorem}
\begin{proof}
We have, for any two languages $L_1, L_2$, $GS(L_1,L_2,R) =
GSCO(GSCO^*(L_1), GSCO^*(L_2))$, where $R$ is the common symbols
between the alphabets of $L_1$ and $L_2$.  By theorems
\ref{l1l2reg}, \ref{reg*l1l2}, we have the result $GS(X,Y,R)$ is
regular.
\end{proof}

Thus, we have found that the class of generalised splicing languages
 $GS (X,Y,R)$ where $X,Y,R$ are as mentioned above, are regular. We
 note that $GS(X,Y,R)=GSCO(GSCO^*(L_1),GSCO^*(L_2)),$ where $R$ is the
 set of  common symbols between the alphabets of $L_1$ and $L_2$.

 This result holds for any set $R'$ (which has only symbols: words of length 1)
  such that $R \subseteq R'$.  The elements in the set $R'-R$ will
  not be the common symbols of the languages in $X$ and $Y$.
 In that case, $GS (X,Y,R')$ will be equal to $GS(X,Y,R)$ since, the
 symbols which are participating in the
 crossover operation not common to the languages , will yield only empty sets.  In a more general
 sense, we have the result that, for  $X,Y \in \{FIN, REG, LIN, CF, CS, RE
 \}, \sg \in FIN$  such that $\sg$ contains only words of length 1,
  the subclass of generalized splicing
 languages namely $GS(X,Y,\sg)$, (which is a subclass of the class
 $GS(X,Y,FIN)$) is always regular.

\section{Conclusion}

In this paper, we have applied an  operation namely $GSCO$ over
languages to study some sub classes of generalised splicing
languages. Using the $GSCO$ operation, it is proved that, for  $X,Y
\in \{FIN, REG, LIN, CF, CS, RE
 \}, \sg \in FIN$  such that $\sg$ contains only words of length 1,
  the subclass of generalized splicing
 languages namely $GS(X,Y,\sg)$, (which is a subclass of the class
 $GS(X,Y,FIN)$) is always regular.

This paper gives a scope for developing the whole theory of
generalised splicing in similar lines to the theory of $H$-system,
which in some sense, is a journey from two dimensions to three
dimensions. The extensive study of generalised splicing can help
both in $H$-system as well as generalised splicing (if $L_1=L_2$ in
generalised splicing, we get back  $H$-system).

This study can be extended to study the other classes of
 generalised splicing languages.

%%%%%%%%%%%%%%%%%%%%%%%%%%%%%%%%%%%%%%%%%%%%%%%%%%%%%%%%%%%%%%%%%%%%%%
%Bibliography file. Do not delete.

\end{document}